\documentclass[preprint,showpacs,preprintnumbers,amsmath,amssymb]{revtex4}

\usepackage{graphicx} 
\usepackage{epsfig} 
\usepackage{mathptmx} 
\usepackage{times} 
\usepackage{amsmath} 
\usepackage{amssymb}  
\usepackage{color}
\usepackage{latexsym}
\usepackage{psfrag}
\usepackage{setspace}

\newcommand{\be} {\begin{equation}}
\newcommand{\ee} {\end{equation}}
\newcommand{\bc} {\begin{center}}
\newcommand{\ec} {\end{center}}

\newcommand{\BEQ}{\begin{equation}}
\newcommand{\EEQ}{\end{equation}}
\newcommand{\BER}{\begin{eqnarray}}
\newcommand{\EER}{\end{eqnarray}}
\newcommand{\BERW}{\begin{eqnarray*}}
\newcommand{\EERW}{\end{eqnarray*}}
\newcommand{\BDM}{\begin{displaymath}}
\newcommand{\EDM}{\end{displaymath}}

\newcommand{\pdd}[2]{\frac{\partial #2}{\partial #1}}
\newcommand{\pddn}[3]{\frac{\partial^{#2} #3}{\partial #1^{#2}}}

\newcommand{\avg}[1]{\left\langle #1 \right\rangle}

\newcommand{\BFig} {\begin{figure}}
\newcommand{\EFig} {\end{figure}}

\def\blfootnote{\xdef\@thefnmark{}\@footnotetext}

\def \llabel#1 {\label{#1} }

\def \ba{{\bf a}}
\def \bF{{\bf F}}

\def \beq{\begin{equation}}
\def \eeq{\end{equation}}
\def \bea{\begin{eqnarray}}
\def \eea{\end{eqnarray}}

\begin{document}

\title{Computational Coarse Graining of a \\Randomly Forced 1-D Burgers Equation}

\author{Sunil Ahuja}
\email{sahuja@princeton.edu}
\affiliation{%
Mechanical and Aerospace Engineering, \\
 Princeton University, Princeton, NJ 08544, USA
}%
\author{Victor Yakhot}%
 \email{vy@bu.edu}
\affiliation{%
Aerospace and Mechanical Engineering, \\
Boston University, Boston, MA 02215, USA
}%
\author{Ioannis G. Kevrekidis}%
 \email{yannis@princeton.edu}
\affiliation{%
Chemical Engineering, PACM, and Mathematics, \\
 Princeton University, Princeton, NJ 08544, USA
}%

\date{\today}

\begin{abstract}
We explore a computational approach to coarse graining the evolution
of the large-scale features of a randomly forced Burgers equation in
one spatial dimension.
The long term evolution of the solution energy spectrum appears
self-similar in time.
We demonstrate coarse projective integration and coarse dynamic
renormalization as tools that accelerate the extraction of
macroscopic information (integration in time,  self-similar shapes,
and nontrivial dynamic exponents) from short bursts of appropriately
initialized direct simulation.
These procedures solve numerically an effective evolution equation
for the energy spectrum without ever deriving this equation in
closed form.
\end{abstract}

\maketitle

\section{Introduction}

The behavior of physical systems is frequently observed, and
modeled, at different levels of complexity.
Laminar Newtonian fluid flow is a case in point: it is modeled at
the atomistic level through molecular dynamics simulators; at a
``mesoscopic" level through Lattice Boltzmann models; and (in
physical and engineering practice) through continuum macroscopic
equations for density, momentum and energy: the Navier--Stokes.
What makes the latter, coarse grained description possible is an
appropriate closure, in this case Newton's law of viscosity, which
{\em accurately models} the effect of higher order, unmodeled
quantities (such as the stresses) on the variables (``observables")
in terms of which the model is written (the velocity gradients).
Such closures are often known from physical and engineering
observation and practice long before their mathematical
justification becomes available (in this case, through
Chapman--Enskog expansions and kinetic theory).
It is, of course, tempting to consider a situation in which the fine
scale description is the Navier--Stokes equations themselves, and
the coarse grained description is an evolution equation for some
observables of turbulent velocity fields, or possibly of ensembles
of such fields.
Discovering the appropriate observables and deriving such evolution
equations has been a long-standing ambition in both physics and
mathematics.
Our goal here is much more modest: we present a very simple
illustration of a problem for which a fine scale description is
available, and for which (based on extended observations of direct
simulation) we have reason to believe that a coarse grained
evolution equation exists -- yet it is not available in closed form.
We illustrate how (with a good guess of the appropriate coarse
grained observables) we {\em circumvent} the derivation of a closed
form coarse grained model, but still perform scientific computing
tasks at the coarse grained level.
The numerical tasks we demonstrate are {\em coarse projective
integration}, which accelerate the (coarse grained) computations of
the system evolution, and {\em coarse dynamic renormalization},
which (when the coarse grained evolution is self-similar, as the
case {\em appears to be} here) targets the computation of the
self-similar solution shape and the corresponding exponents.
This is an illustration of the so-called Equation-Free framework for
coarse grained scientific computation in the absence of {\em
explicit} quantitative closures and the resulting coarse grained
evolution equations~\cite{manifesto, pnas, manifesto_short}.
We consider the one dimensional in space, randomly forced Burgers
equation,
\beq \pdd{t}{u} + \frac{1}{2} \pdd{x}{u^2} = \nu_{\mbox{\small hyp}}
(-1)^{n+1} \pddn{x}{2n}{u} + f(x,t), \label{burgers} \eeq
subject to periodic boundary conditions over the domain~$x \in [0,
2\pi]$.
We start with random {\em very low energy} initial conditions.
We are interested in the relatively long term and large scale
properties of this system in the inviscid limit, when subject to a
random forcing at the small scales.
In order to achieve this inviscid limit numerically, we consider a
general dissipation term of the form~$\nu_{\mbox{hyp}} \, (-1)^{n+1}
\partial^{2n} u / \partial {x}^{2n}$ following~\cite{chekhlov95}.
The coefficients~$\nu_{\mbox{hyp}}$ and~$n$ are chosen so that the
dissipation acts only at the highest wavenumbers; here, their values
are~$\nu_{\mbox{hyp}} = 10^{-54}$ and~$n = 7$.
This choice of dissipation is based on the assumption that the
universal ranges of the system are not sensitive to a particular
choice of the parameters~$\nu_{\mbox{hyp}}$
and~$n$~\cite{chekhlov95}.
The white--in--time random force~$f(x,t)$ is defined in Fourier
space by
\bea \overline{ f(k,\omega)f(k',\omega')} = {\cal P}
\exp\Big[-\frac{(k - k_s)^2} {\sigma_f^2} \Big]
\delta(k-k')\delta(\omega-\omega'), \qquad \mbox{with } {\cal P} =
\frac{f_0^2}{2\pi \sigma_f^2}, \label{forcing} \eea
where~$k$ and~$\omega$ are the spatial and temporal frequencies.
The forcing term has a peak at wavenumber~$k=k_s=1000$ and dies off
at large wavenumbers; the dissipation term is essentially zero up to
around wavenumber~$k = 3000$ and only becomes important at much
larger wavenumbers; see~Fig.~\ref{fig:force}.
%
\BFig \bc
\psfrag{k}{$k$}
\psfrag{forcing}{\small forcing}
\psfrag{dissipation}{\small dissipation}
\includegraphics[scale=0.4]{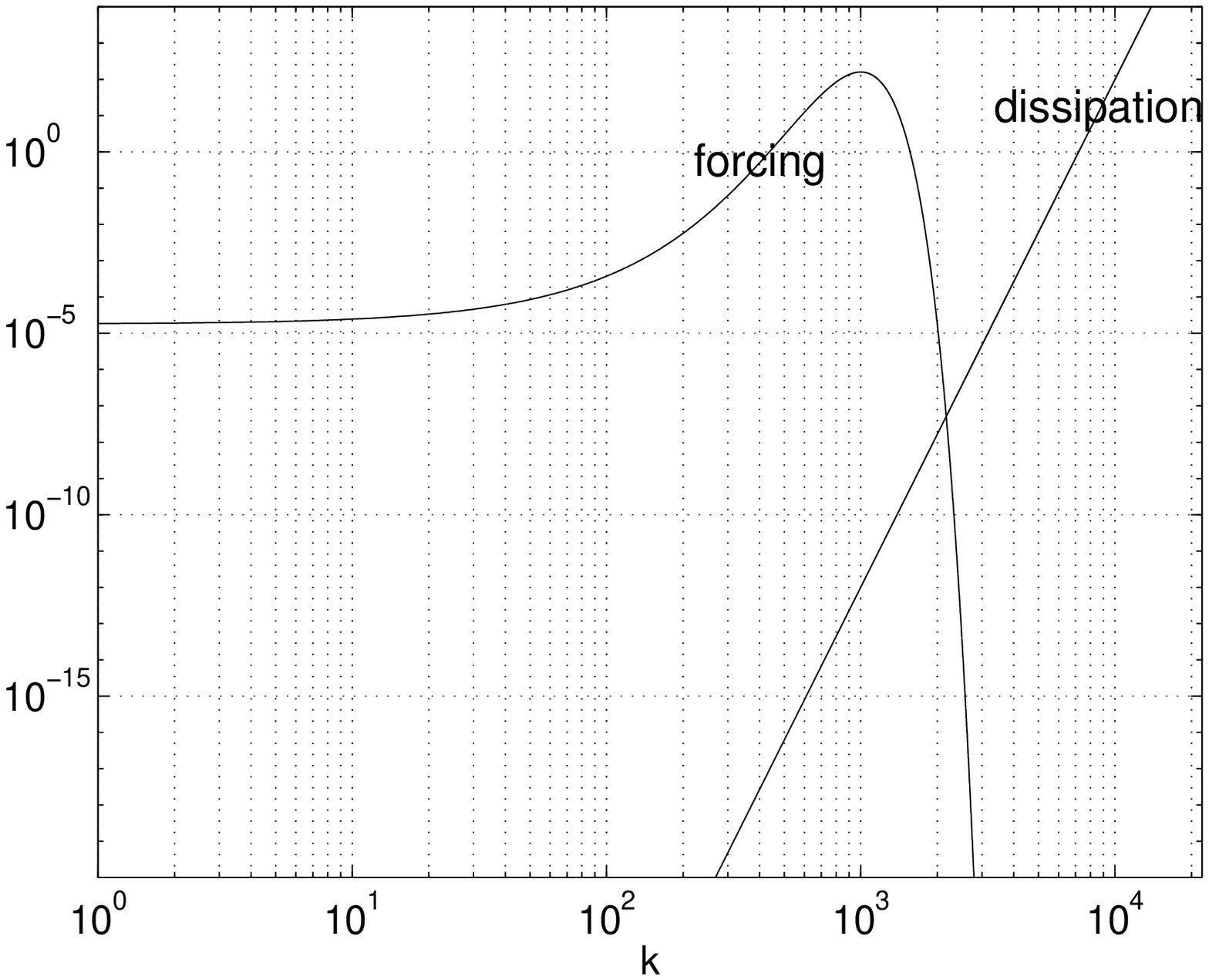}\\
\caption{Forcing amplitude~$\frac{f_0^2}{2\pi\sigma_f}
\exp[-\frac{(k - k_s)^2}{\sigma_f^2} ] $ and dissipation
term~$\nu_{\mbox{hyp}} k^{2n}$.} \label{fig:force} \ec \EFig

Before we start our computations, we briefly recall certain known
features of the equations and their dynamics.
It is quite remarkable that equations~(\ref{burgers},\ref{forcing})
have non-trivial asymptotic behavior in both the
ultraviolet~(UV,~$k>k_{s}$) and the infrared~(IR,~$k<k_{s}$) limits.
Attempts to use~(\ref{burgers},\ref{forcing}) as a simplified
1--D~model for the investigation of small scale features of
3--D~turbulence, aimed at recovering~$E(k) \propto k^{-5/3}$ in the
small scale limit, were based on deep similarities between the two:
the total energy in this system is $\overline{u^{2}}/2\equiv
K\approx ({\cal P}L)^{2/3}$  and, as in 3--D~turbulence, the
inertial range~$k_s\ll k\ll k_{d}$ where~$k_{d}$~is the dissipation
wave number, is characterized by a constant energy flux~$J_{E}={\cal
P}$, which, in the interval~$r\ll L=1/k_s$, is reflected in the
analytic behavior of the third-order structure function
\begin{equation}
S_{3}(r)=\overline{(\delta_{r} u)^{3}}\equiv
\overline{(u(x+r)-u(x))^{3}}=-12{\cal P}r.
\end{equation}
%
%
Even in decaying high Re--turbulence this equation is correct:
$\partial E/ \partial t $ is very small.
At first glance, due to the shock formation leading to the
$E(k)\propto k^{-2}$~energy spectrum (rather than~$E(k)\propto
k^{-x}$ with the exponent~$x\approx 5/3$), these attempts failed.
However, the model revealed a non-trivial bi-scaling behavior of the
structure functions~$S_{n}(r)=\overline{(u(x+r)-u(x))^{n}}\propto
r^{n}$ for~$n\leq 1$ and $S_{n}(r)\propto r$~for~$n\geq
1$~\cite{frisch}.
The model also generated asymmetric probability densities with
algebraically decaying tails~$P(\delta_{r}u)$ for large amplitude
fluctuations~$\delta_{r}u<0$~\cite{chekhlov95,vyPRL96}.
The properties of this tail attracted the attention of various
groups and are reasonably well understood in terms of the
dissipation anomaly introduced by~\cite{polyakov95}; they have also
been analytically obtained in~\cite{e97,e99,bec00}.
The Burgers equation stirred by a force with the~$k^{-1}$~spectrum
does lead to the logarithmically corrected Kolmogorov energy
spectrum discovered in~\cite{chekhlov95b} and does not have
multifractality~\cite{mitraArXiv}.

The IR properties of~(\ref{burgers},\ref{forcing}) in the
range~$k\ll k_s$, on which we focus in this paper, are also quite
interesting and non-trivial.
It was determined, on the basis of the one-loop calculation (also
numerically verified in~\cite{vyPRL88}) that in the limit~$k\ll
k_s$, (\ref{burgers},\ref{forcing})~is equivalent to the Burgers
equation driven by a random force with~$\overline{f(k)^{2}}\propto
k^{2}$--spectral density, which is called the KPZ equation,
governing various interfacial phenomena~\cite{kpz}.
The renormalization group methodology applied to the KPZ
equation~\cite{forster77} led to the equilibrium energy
spectrum~$E(k)\propto k^{D-1}=\mbox{const}$~(here~$D=1$) and to a
non-Gaussian {\it dynamic} fixed point with non-trivial dynamic
scaling exponents,
\begin{equation}
\omega\propto k^{\frac{3}{2}}, \label{exp}
\end{equation}
indicating strongly non-diffusive behavior:~$r(t)\propto
t^{\frac{3}{2}}$. These predictions, including dynamic scaling, have
been numerically verified in~\cite{vyPRL88}.
Numerical simulations in the IR range become progressively slower;
our computer-assisted approach aims at accelerating such simulations
(see also~\cite{kesslerPRE}).

Our numerical simulator integrates equation~(\ref{burgers}) in time
using a pseudo-spectral scheme with a~$N=22000$~Fourier mode spatial
discretization.
Thus, the smallest scale resolved by our simulator is~$\Delta x = \pi/N = 
1.428 \times 10^{-4}$. 
The nonlinear term is computed using the fast Fourier transform.
The temporal discretization treats the linear term exactly and uses
a third order Runge--Kutta--Wray scheme for the nonlinear term.
The random forcing term in Fourier space is~$f(k,t) = \sqrt{{\cal
P}/ \Delta t} \sigma_k$, where $\sigma_k$~is a random number
generated from a Gaussian distribution with zero mean and unit
standard deviation, and $\Delta t$~is the time step of integration.
The various parameters chosen are~$\Delta t = 5 \times 10^{-6}$,
$f_0 = 5 \times 10^2$, $k_s = 1000$, $\sigma_f = 250$,
$\nu_{\mbox{hyp}} = 10^{-54}$, and~$n = 7$.
An ensemble of 16~distinct realizations was computed, and the
results averaged to achieve good statistics.

We begin by observing direct numerical simulations of the evolution
of equation~(\ref{burgers}) over time.
Fig.~\ref{fig:init_evl} shows such a transient, starting from the
zero initial condition,~$u(k,t=0) = 0$, for all~$k$.
The energy spectra~$E(k,t)$ are plotted after~$1000$, $5000$,
$25000$, and~$300000$ time steps; the insets show the corresponding
1D~velocity fields.
Visually, we can discern no clear trend in the evolution of the
velocity fields themselves, which appear random; yet, we easily see
a definite smooth structure in the evolution of the energy spectrum.

For~$k\geq 700$, the spectrum seems to quickly achieve a steady
state profile within less than 1000~time steps: the energy provided
by the noise is balanced by the strong dissipation at high
wavenumbers.
It is much more interesting to observe~$E(k)$ for~$1 \leq k \leq
500$; here the spectrum quickly acquires a ``corner-like" shape
which appears to evolve smoothly in time, and moves towards lower
wavenumbers while maintaining this shape.
This ``traveling front-like"  evolution progressively slows down as
the ``corner"  evolves towards lower wavenumbers.
\BFig \bc
\psfrag{a}{(a)}
\psfrag{b}{(b)}
\psfrag{c}{(c)}
\psfrag{d}{(d)}
\psfrag{e}{$E(k)$}
\psfrag{k}{$k$}
\psfrag{g}{$k_f$}
\includegraphics[scale=0.75]{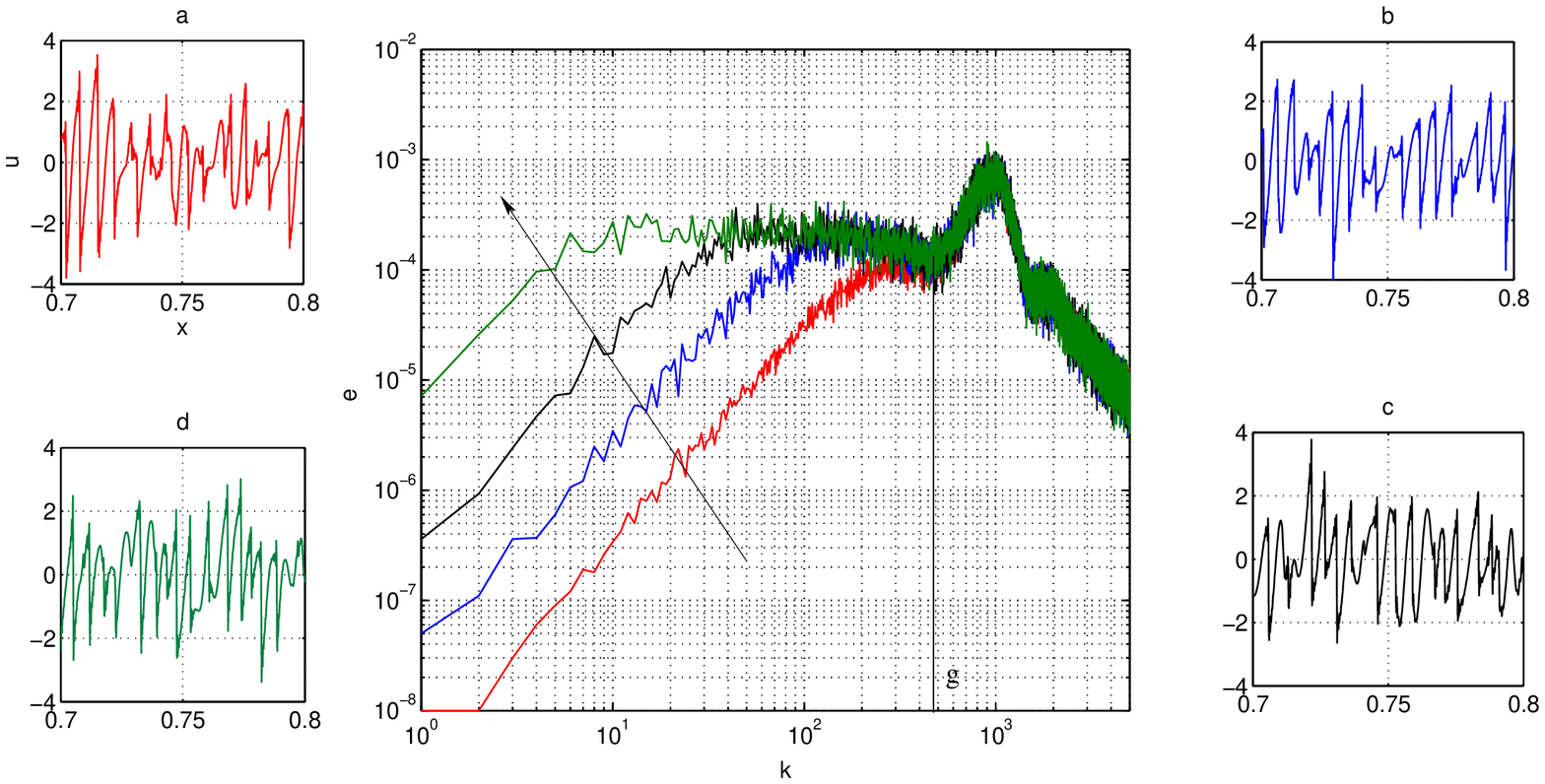}
\caption{Evolution of $E(k)$ from the zero initial condition $u(k,0)
= 0$, for all~$k$.
The spectra are shown after~1000, 5000, 25000, and 300000~time
steps, with time increasing in the direction of the arrow.
The velocity fields corresponding to the spectra are shown in the
insets, in the same color as the spectra.}
\label{fig:init_evl} \ec
\EFig

Theoretical considerations~\cite{forster77,vyPRL88} indicate that in
the range~$k<k_s$, the evolving energy spectrum~$E(k)$ consists of
three regimes.
For~$k=O(k_s)$, the spectrum is not universal and depends on the
properties of the forcing function.
For~$1/L\ll k_{I}(t)\ll k\ll k_s$,~$E(k)=k^{D-1}=
\mbox{const}$~(here~$D=1$); and in the time-dependent large-scale
limit~$1/L\ll k<k_{I}(t)$, the energy spectrum is given by a
universal relation~$E(k)\propto k^{D-1+2}\propto k^{2}$, which we
observe in our simulations.
Superficially, the temporal evolution we observe resembles the
time-dependence of a two dimensional flow driven by a
small~($k_s=O(1)$) random force, eventually leading to the energy
spectrum piling up on the smallest (integral) wave
number~$k_{i}=\pi/L$ and leading to the IR asymptotics~$E(k)\propto
k^{-3}$.
We stress that the situation we are interested in may be quite
different: in 2D, due to enstrophy conservation in the infrared
range~($k<k_s$), there exists a constant, {\it nonzero}, energy flux
toward large scales, leading to the so called ``inverse cascade".
The third-order structure function in a 2D flow
is~$S_{3}=\frac{3}{2}{\cal P}r\neq 0$.
In the one dimensional flow we are interested in here, the large
scale thermodynamic equilibrium is flux-free and $S_{3}$~must be
equal to zero, provided the system is large enough,~that
is,~$k_{I}/k_s\ll 1$.

We do not work at this very long time regime; we rather observe the
evolution at times {\em intermediate} between this regime and the
(relatively) short time it takes for the spectrum to stabilize
for~$k > 700$.
This is the regime in which smooth and, as we argue below,
apparently self-similar evolution of the energy spectrum prevails.

Based on the study of many transients like those in
Fig.~\ref{fig:init_evl}, we decided to observe the system dynamics
in terms of the energy spectrum~$E(k,t)$ only, which is defined as
\beq E(k,t) = \avg{\hat{u}(k,t) \hat{u}^\ast(k,t)}, \eeq
where $\hat{u}(k,t)$~is the spatial Fourier transform of the
velocity field~$u(x,t)$, and $\avg{\cdot}$~denotes an ensemble
average~(here over the 16~realizations).
A formal diagrammatic expansion for the evolution of the energy
spectrum corresponding to~(\ref{burgers},\ref{forcing}) can be
developed~\cite{wyld61}.
In the limit~$k\rightarrow 0$, taking into account that all
odd-order correlation functions rapidly tend to zero, the solution
can be accurately represented in terms of an infinite series
involving complicated integral expressions with kernels which are
convolutions of the energy spectrum~$E(k)$ only.
Based on this, we assume that an evolution equation of the general
form
\beq \pdd{t}{E(k,t)} = \mathcal{L}[E(k,t)] \eeq
exists and closes for the (ensemble averaged)~$E(k,t)$.
We do not know the nature of the operator~$\mathcal{L}$ -- it may
not be local (in~$k$--space), and the equation may not be a partial
differential equation -- yet the implication is that
knowing~$E(k,0)$ is enough to predict the (realization ensemble
averaged)~$E(k,t)$ for later times.
If an evolution equation closes with respect to the expected
spectrum {\em only}, then the remaining degrees of freedom (the
phases of the velocity field Fourier coefficients) do not explicitly
appear in this equation.
This implies that either these variables {\em do not couple} to the
spectrum evolution, or that their expected dynamics become quickly
slaved to the spectrum evolution (with a relaxation time that is
small compared to the spectrum evolution time scales).
We test this latter ansatz by observing the solution ensemble not
only through the spectrum~$E(k,t)$ but also through the evolving
third order structure function
\beq S_3(r,t) = \avg{\overline{(u(x,t) - u(x+r,t))^3}}. \eeq

\BFig \bc
(a) \hspace{3in} (b) \\
\includegraphics[scale=0.5]{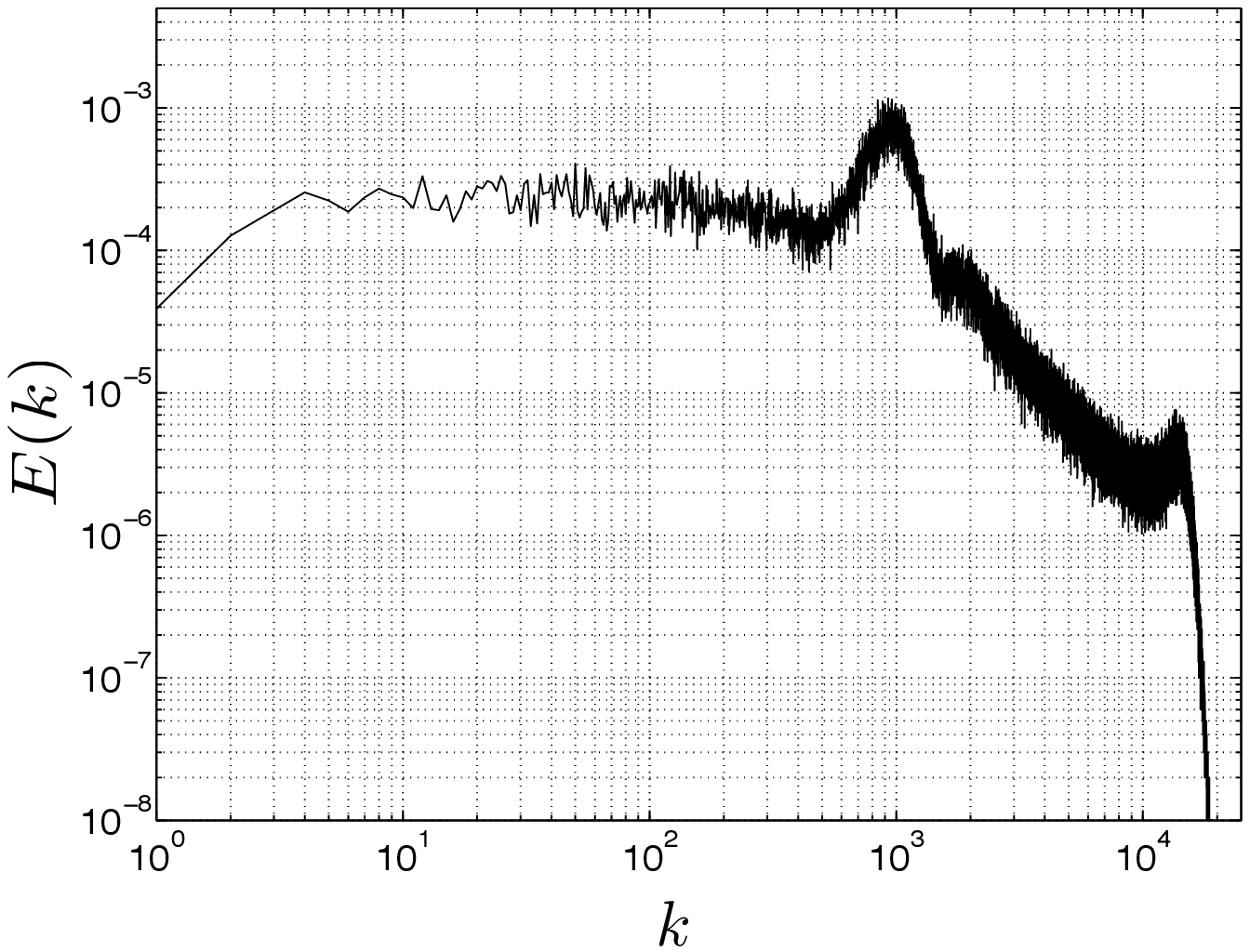}
\hspace{0.5cm}
\includegraphics[scale=0.5]{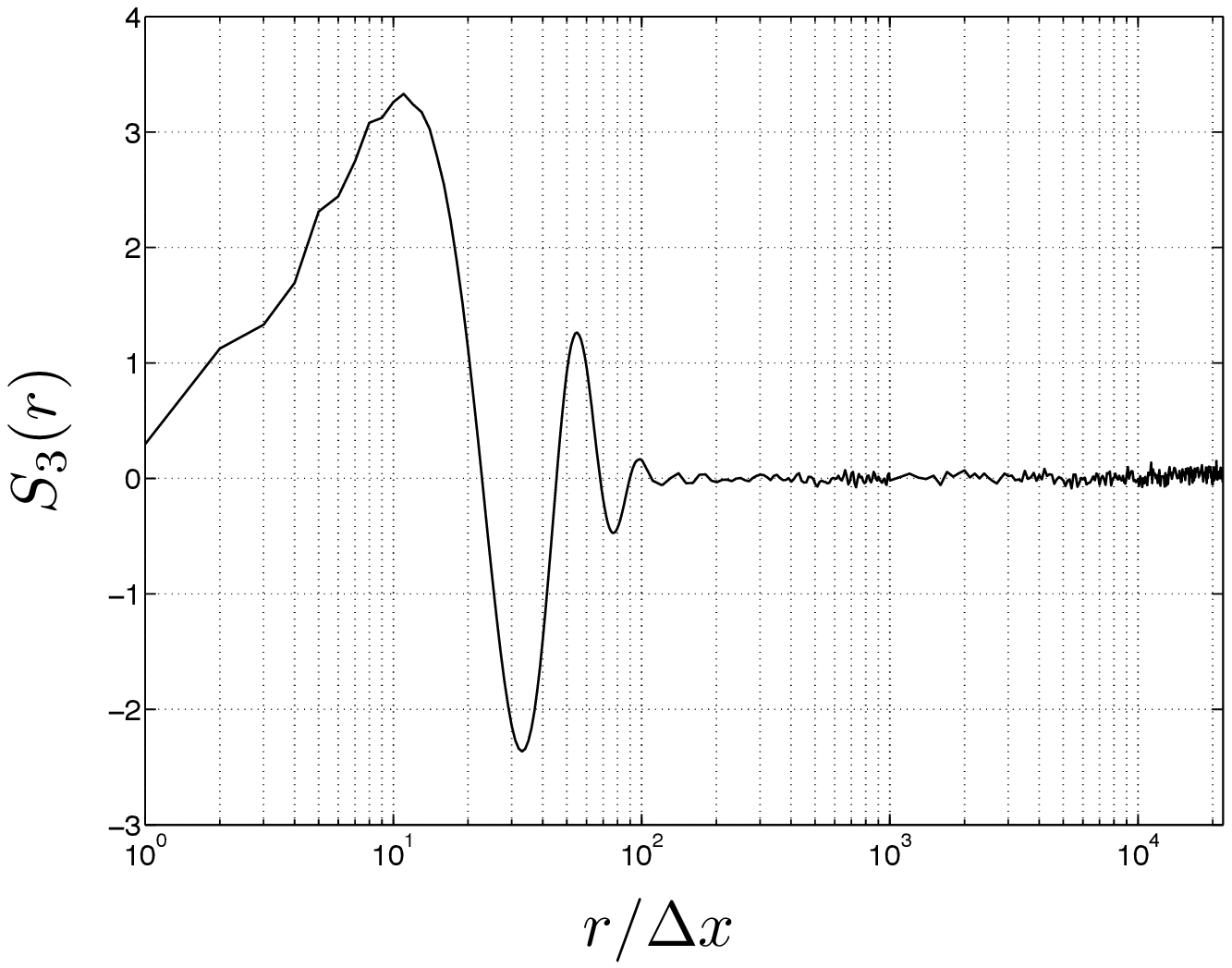}
\caption{Long term solution, after~$1.25 \times 10^6$ time steps
(a)~$E(k)$ vs.~$k$; (b)~$S_3(r)$ vs.~$r/ \Delta x$, 
where~$\Delta x = \pi / N$ is the smallest scale resolved in 
our simulations. }
\label{fig:long_term} \ec \EFig

For completeness, the ``extremely    long term" steady state
of~$E(k,t)$ and~$S_3(r,t)$, obtained after integrating for~$1.25
\times 10^6$ time steps is included in Fig.~\ref{fig:long_term}.
Fig.~\ref{fig:long_term}(a) suggests that $E(k)$~reaches a constant
value of~$2 \times 10^{-4}$ in the range~$1 \leq k \leq 150$.
The structure function~$S_3(r)$, plotted in
Fig.~\ref{fig:long_term}(b), is practically zero in the range~$200
\Delta x \leq r \leq 2 \pi$.
The steady state value of~$S_3(r)$ is nonzero for~$r<200 \Delta x$,
due to the force acting primarily at small scales (corresponding
to~$k=1000$) and strong dissipation at the smallest scales.

The remainder of the paper is organized as follows:
In Section~\ref{section:selfsim} we discuss simulations suggesting
an apparent dynamic self-similarity in the intermediate-time
evolution of the spectrum.
Section~\ref{section:eqnfree} briefly outlines equation free
computational techniques, and demonstrates coarse projective
integration of the unavailable equation for the expected spectrum
evolution.
Section~\ref{section:renorm} then shows how to implement a dynamic
renormalization fixed point algorithm that converges on the
self-similar shape and pinpoints the temporal similarity exponent of
its evolution.
We conclude with a brief discussion.
\section{Apparent Dynamic Self-similarity}
\label{section:selfsim}
Here we argue that the transient evolution of~$E(k)$ exhibits
dynamic self-similarity.
Fig.~\ref{fig:cartoon1} clearly shows that the qualitative shape of
the spectrum in the region~$1\leq k \leq 500$ remains effectively
unchanged, while gradually ``travelling" to the left, slowly filling
up the lower wavenumbers.
We make this observation more precise as follows:
We define a ``fulcrum point"~$k=k_f$ as the leftmost end of the
apparently steady region of~$E(k)$; here~$k_f \approx 450$.
As shown in Fig.~\ref{fig:cartoon1}(a), we draw straight line
segments of various slopes~$m_j$ (logarithmic in~$E$ as well as
in~$k$) starting at the fulcrum and intersecting successive computed
spectra.
Let~$d_{m_j}(t)$~be the distance of the point of intersection of
each such line with the spectrum at time~$t$.
Our ansatz is that the distance~$d_{m_j}(t)$ becomes uniformly
stretched in (logarithmic)~time.
This is tested as follows: let the stretching factor along a
line of slope~$m_j$ be~$r_{m_j}(t) = d_{m_j}(t)/d_{m_j}(t_1)$, where
$t_1$~is a reference time.
Fig.~\ref{fig:cartoon1}(b) is a plot of~$r_{m_j}$~versus~$m_j$ at
various times in the order listed in the figure caption; it strongly
suggests that the spectrum gets stretched by the same amount along
its entire profile (notice that no intersections arise with~$m_j
\lesssim -0.1$).
Fig.~\ref{fig:cartoon1}(c) shows the time evolution of the
stretching factor~$r(t)$ (averaged over the lines of different
slopes), clearly showing linearity in logarithmic time.
Thus, the transient evolution of~$E(k)$ appears dynamically
self-similar: the spectrum in the wavenumber region~$1 \leq k
\lesssim 450$ becomes uniformly stretched in logarithmic time, while
maintaining its shape.
We take advantage of this information in speeding up
computations in the next section.
\BFig \bc
\psfrag{a}{$d_m(t_1)$}
\psfrag{b}{$d_m(t_n)$}
\psfrag{c}{fulcrum}
\psfrag{d}{line of slope $m$}
\psfrag{e}{$\log \, E(k)$}
\psfrag{f}{(a)}
\psfrag{g}{$k_f$}
\psfrag{i}{{\small slope = 0.48}}
\psfrag{k}{$\log \, k$}
\psfrag{p}{{\small slope}, $m$}
\psfrag{q}{ {\em \small t}}
\psfrag{r}{{\small mean stretch}}
\psfrag{t}{$\log t$}
\psfrag{u}{{\small stretch}, $r_m$}
\psfrag{y}{(b)}
\psfrag{z}{(c)}
(a) \hspace{4.5cm} (b) \hspace{4.5cm} (c) \\
\includegraphics[scale=0.35]{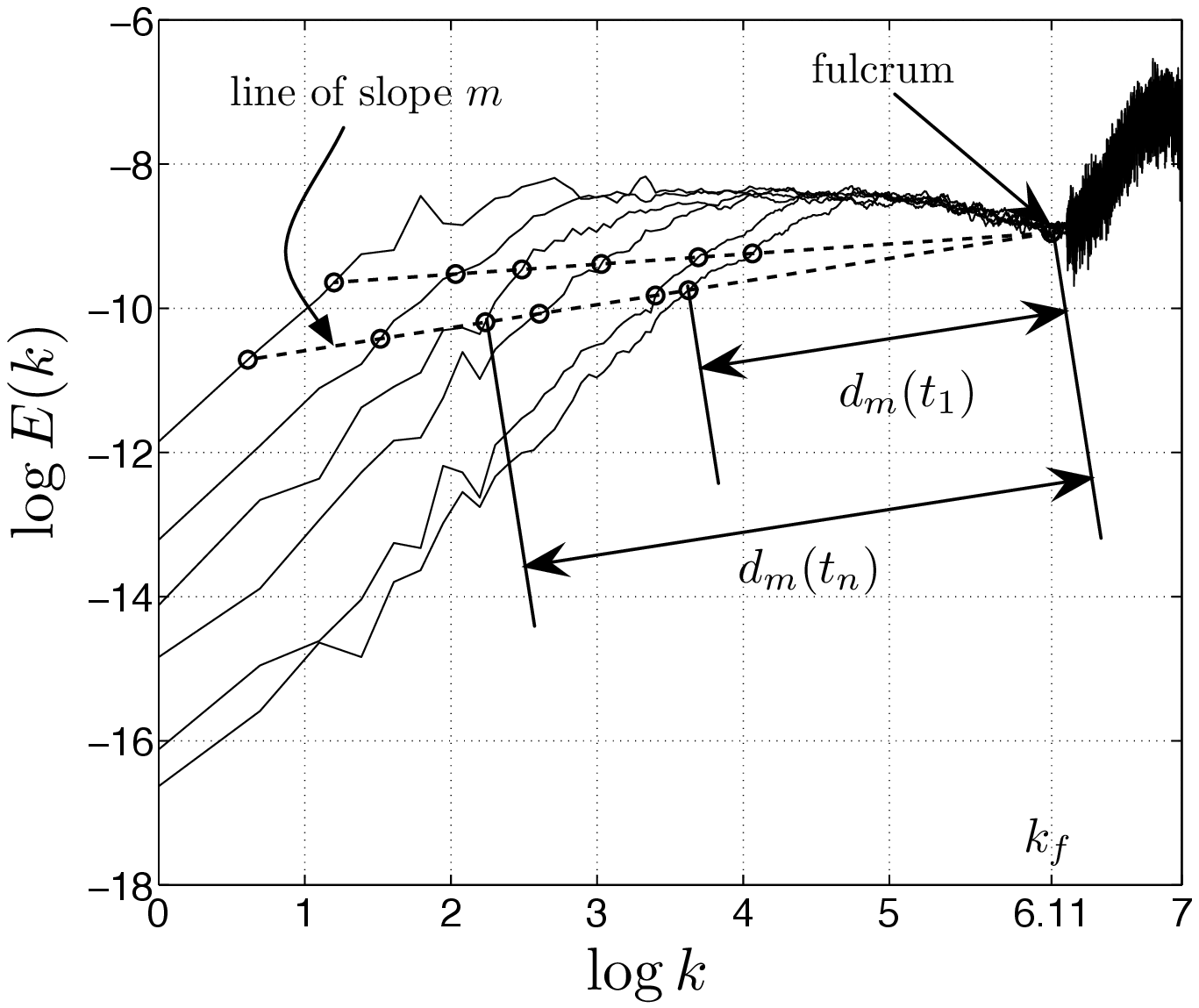}
\includegraphics[scale=0.35]{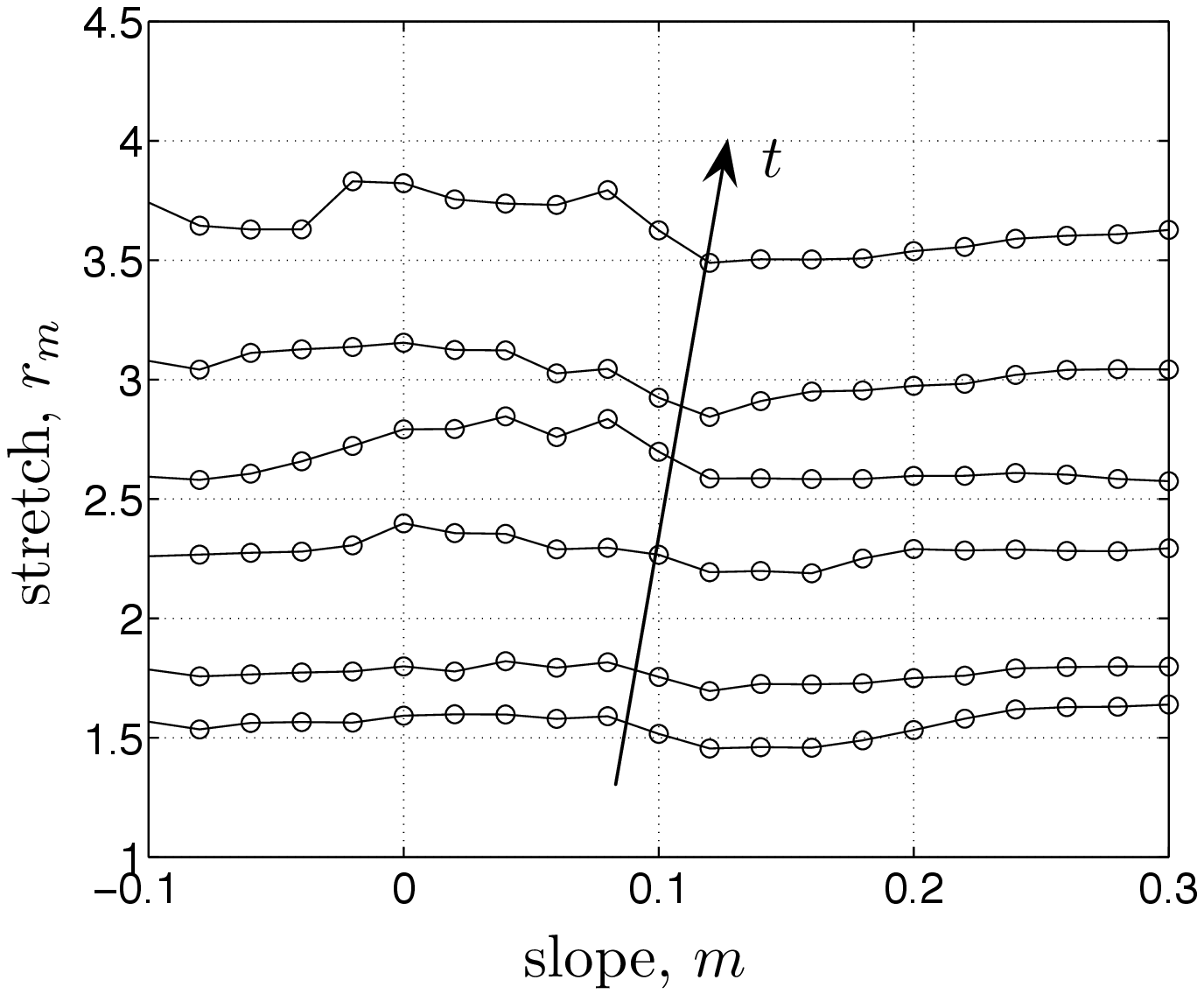}
\includegraphics[scale=0.35]{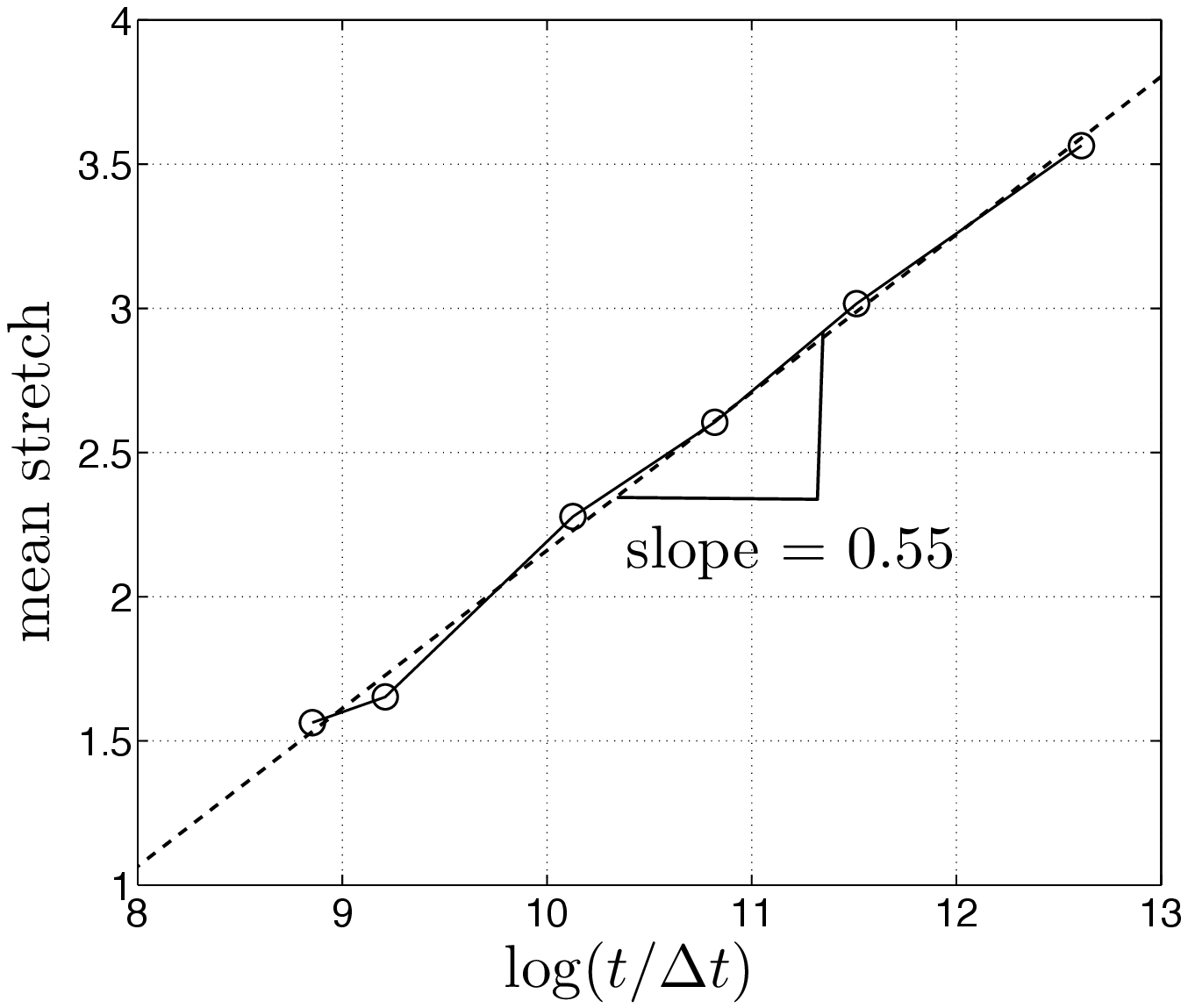}
\caption{(a)~Schematic diagram showing calculation of the stretching
factor. For a line of slope~$m$, $r_m(t_n) = d_m(t_n)/d_m(t_1)$.
(b)~Stretching factor~$r_m(t_i)$ vs. slope~$m$ after
$7,10,25,50,100, \mbox{ and } 300 \, (\times 1000)$ time steps; the
direction of time is indicated by the arrow.
Stretching is measured with respect to the spectrum at $t=2500
\Delta t$. (c)~Mean stretch~$r(t)$ vs.~$\log(t/\Delta t)$ (solid
line) and a least squares fit (dotted line).} \label{fig:cartoon1}
\ec \EFig

\section{Coarse Projective Integration}
\label{section:eqnfree}

Based on our simulation observations above, we now hypothesize
that an evolution equation {\em exists and closes} at the level of
the energy spectrum~$E(k,t)$ averaged over our ensemble of
realizations.
If this equation were explicitly available, a typical numerical
integration scheme would start by discretizing it in~$k$ (for
example through finite differences,~FD, or finite elements,~FE) and
then proceed to the derivation of ordinary differential
equations~(ODEs) in time for the vector of values~$\ba$ at the
FD~discretization points, or of the FE~coefficients.
These ODEs would be of the general form~$d\ba/dt = \bF(\ba)$; the
temporal evolution would be obtained using established numerical
integration techniques, of which the simplest is arguably the
explicit forward Euler method
\beq
\ba(t_{0} + \Delta t) = \ba(t_{0}) + \Delta t \, \bF(\ba(t_0)).
\eeq
Clearly, implementing this, as well as other initial value solvers,
requires an evaluation of the time derivative (the right hand side
of the ODE set) at prescribed system states.
In projective integration~\cite{gk1,gk2} this time derivative is not
obtained through such a function evaluation of~$\bF(\ba)$, since
$\bF(\ba)$~is not available in closed form; instead, it is obtained
from processing the results of short bursts of simulation: starting
at time~$t_0$, we integrate the original equation for a short time
with a time step~$\delta t \ll \Delta t$ to obtain a sequence
$\ba(t_0), \ba(t_0 + \delta t), \ba(t_0 + 2 \delta t),\, \ldots, \,
\ba(t_0 + n \delta t)$, where~$n \delta t < \Delta t$.
We can then use, for example, the last two points of this sequence
to estimate the time derivative, and predict~$\ba(t_{0} + \Delta t)$
as
\beq
\ba(t_{0} + \Delta t) \approx \ba(t_0 + (n-1) \delta t) +
\frac{\Delta t - (n-1)\delta t}{\delta t} \big( \ba(t_0 + n \delta
t) - \ba(t_0 + (n-1) \delta t)\big ) ;
\eeq
better estimators (least squares, maximum likelihood) of the time
deerivative can clearly be used as well.
The procedure provides an approximation of the ``projected in time"
state~$\ba(t_0+\Delta t)$, which can then be used to restart the
computation, and the process repeats.

When the only available tool is a ``black box" simulator of the
system evolution equation, projective integration may significantly
accelerate computations for systems with gaps in their eigenvalue
spectra (separation of time scales).
However, in our case, the equation we want to projectively integrate
is an equation for {\em energy spectrum evolution}, while the
available simulator is a simulator for {\em velocity fields}; we
want to projectively integrate the {\em coarse grained behavior}
(spectra) through observations of the {\em fine scale behavior}
(velocities).
This now becomes {\em coarse} projective integration.
From our simulation we have available the full state (the velocity
field ensembles) at time~$t_0$; full direct simulation will then
provide full states at each of the intermediate time steps.
We {\em restrict} the full state to the coarse observables (the
discretization of the spectrum); we use these observations to
estimate the time derivative of these coarse observables; and we
finally pass the time derivative estimates to our integrator of
choice (here a simple forward Euler) in order to produce our
approximation of the coarse observables at the next (projected)
point in time~$t_0 + \Delta t$.

Here comes a vital step in our equation-free coarse graining
approach: our ``black box" simulator requires a fine scale
initialization (detailed velocities); yet we only have coarse
grained initial conditions (ensemble-averaged spectrum
discretizations).
For the procedure to continue, we need to somehow construct full
fine-scale states {\em consistent with} the desired coarse initial
conditions.
This construction is the so-called {\em lifting}
step~\cite{pnas,gkt02,manifesto,manifesto_short}.
The missing degrees of freedom in our construction are the {\em
phases} of the velocity fields.
Our assumption that an equation exists and closes in terms of only
ensemble averaged spectra implies that these remaining degrees of
freedom either do not couple with the evolution of the coarse
grained variables we retain, or, if they do (as we expect), they
quickly get slaved to the coarse variables.
``Quickly" here implies a comparison with the time scales of the
evolution of the retained coarse variables themselves.
Exploiting this ``quick slaving" (which is implicit in the
assumption that an equation for the retained coarse variables exists
and closes) is the main point of equation-free computation: since
the fine scale code is available, running it for a short time, even
with wrongly initialized ``additional variables", should provide the
required slaving on demand.
This can be considered a subcase of optimal
prediction~\cite{chorin}.
Given an energy spectrum initial condition, we therefore initialize
the velocity fields as follows: the amplitudes of the velocity
Fourier modes are chosen consistently with the prescribed initial
energy spectrum, while the phases are chosen randomly.
We then let our direct simulator evolve for some initial ``healing"
interval (described below) before we start observing the solution
energy spectrum, in order to estimate its time derivative; this
initial period is allowed so that our ``wrong" phase initialization
becomes ``healed", that is, so that our phases or their statistics
become slaved to the evolution of the spectra.
We monitor this slaving through observations
of the structure function~$S_3(r,t)$ defined earlier.

Here is a concise description of our coarse (spectrum level)
projective (forward Euler) integration.
The procedure starts after we have already simulated (starting with
zero initial conditions) for a period of approximately 3000~time
steps; during this period the ``high--$k$" end of the
spectrum~$E(k>k_f)$ approaches stationarity.

\begin{enumerate}
\item
Starting with a ``current" velocity field initial condition, we
evolve the randomly forced Burgers simulator for a short
interval~$T$, and {\em restrict} the velocity fields at successive
instances within this interval to ensemble-averaged energy spectra.
Fig.~\ref{fig:strch6}(a) shows the restricted spectra at
times~$t/\Delta t =$~3500, 4500, 5500, and~7000.
\item
Using the procedure described in section~\ref{section:selfsim}, we
compute the stretching factors~$r_{m_j}(t)$ along various straight
lines passing through the fulcrum and intersecting these spectra.
Our discretization of the spectrum does not come in the
form~$E(k_i)$ for a number of mesh points~$k_i$ (or~$\log k_i$), but
rather in the form~$(E(m_j),\, k(m_j))$ for a number of
``discretization slopes"~$m_j$; the ensemble-averaged spectrum
evolution appears smooth enough, so that approximately 20~such
discretization angles and simple interpolation is sufficient.
We now compute the rate of change of the stretching factors in
logarithmic time, and use the same to {\em project} the energy
spectrum to a future time~$\gg T$.
Here, the stretching factors~$r_{m_j}$ are computed along straight
lines of slopes~$-0.2<m_j<0.1$, and their rate of change in
logarithmic time is computed using a least squares fit as
\beq r_{m_j}(t) \approx P_1^{m_j} + P_2^{m_j} \log t,
\label{project} \eeq
where~$P_1^{m_j}$ and~$P_2^{m_j}$ are the coefficients of the least
squares fit.
We then use equation~(\ref{project}) to project the stretching
factor to a time~$t^{\ast}\gg T$, and compute the projected spectrum
using
\begin{align}
\log[ E(m_j,\, t^\ast)] &= r_{m_j}(t^\ast) \log[ E(m_j,\, t_1)], \nonumber \\
\mbox{and} \quad \log[ k(m_j,\, t^\ast)] &= r_{m_j}(t^\ast)
\log[ k(m_j,\, t_1)].
\end{align}
An estimate of the spectrum at~$t^\ast = 30000 \Delta t$ using those
at times~$t/\Delta t =$~3500, 4500, 5500, and~7000 is shown in
Fig.~\ref{fig:strch6}(a).
\item
We now {\em lift} the projected spectrum to an ensemble of
consistent velocity fields in order to repeat the process.
This is done by {\em randomizing the phases} for all the
wavenumbers, while retaining the amplitudes corresponding to the
projected spectrum.
More precisely, we generate 16 different initial velocity fields
as~$u(k,t) = \sqrt{E(k,t)} \exp(i\phi_k)$, where~$\phi_k$ is a
random number between $0$~and~$2\pi$ chosen from a uniform
distribution.
\item
We now repeat the process: we evolve the fine scale simulator (the
randomly forced Burgers) with the initial conditions thus obtained;
but before we start using the resulting spectra to estimate a {\em
new} coarse time derivative and make a {\em new} forward Euler
projection, we {\em test} the slaving of the injected degrees of
freedom (the random phases) to our coarse observables (the
spectrum).
This slaving is monitored through the evolution of the newly
initialized structure function~$S_3(r)$.
Fig.~\ref{fig:strch6}(b) shows the structure function from a
simulation that has been running for some time; if we restrict to
energy spectra, and then randomize the phases (that is, lift back to
velocity fields) this phase information is lost.
When we restart the simulation from the lifted velocity fields,
however, we clearly observe that the phases ``heal", that is,
$S_3$~acquires its original shape {\em quickly} compared to the
evolution time scales of the spectrum (in less than 500~time steps,
which is much smaller than the projection horizon of $> 20000$~time
steps).
We continue now the ``short burst" of fine scale simulation to
collect data (restrictions to spectra) for the next projection step;
the procedure~(steps 1--3) then repeats.
\end{enumerate}
\BFig \bc
\psfrag{z}{\small $t=30000$}
\psfrag{r}{$r$}
\psfrag{S3(r)}{$S_3(r)$}
\psfrag{log(k)}{$\log k$}
\psfrag{log[E(k)]}{$\log E(k)$}
\psfrag{t=30400}{$t = 30400$}
(a) \hspace{3in} (b) \\
\includegraphics[scale=0.5]{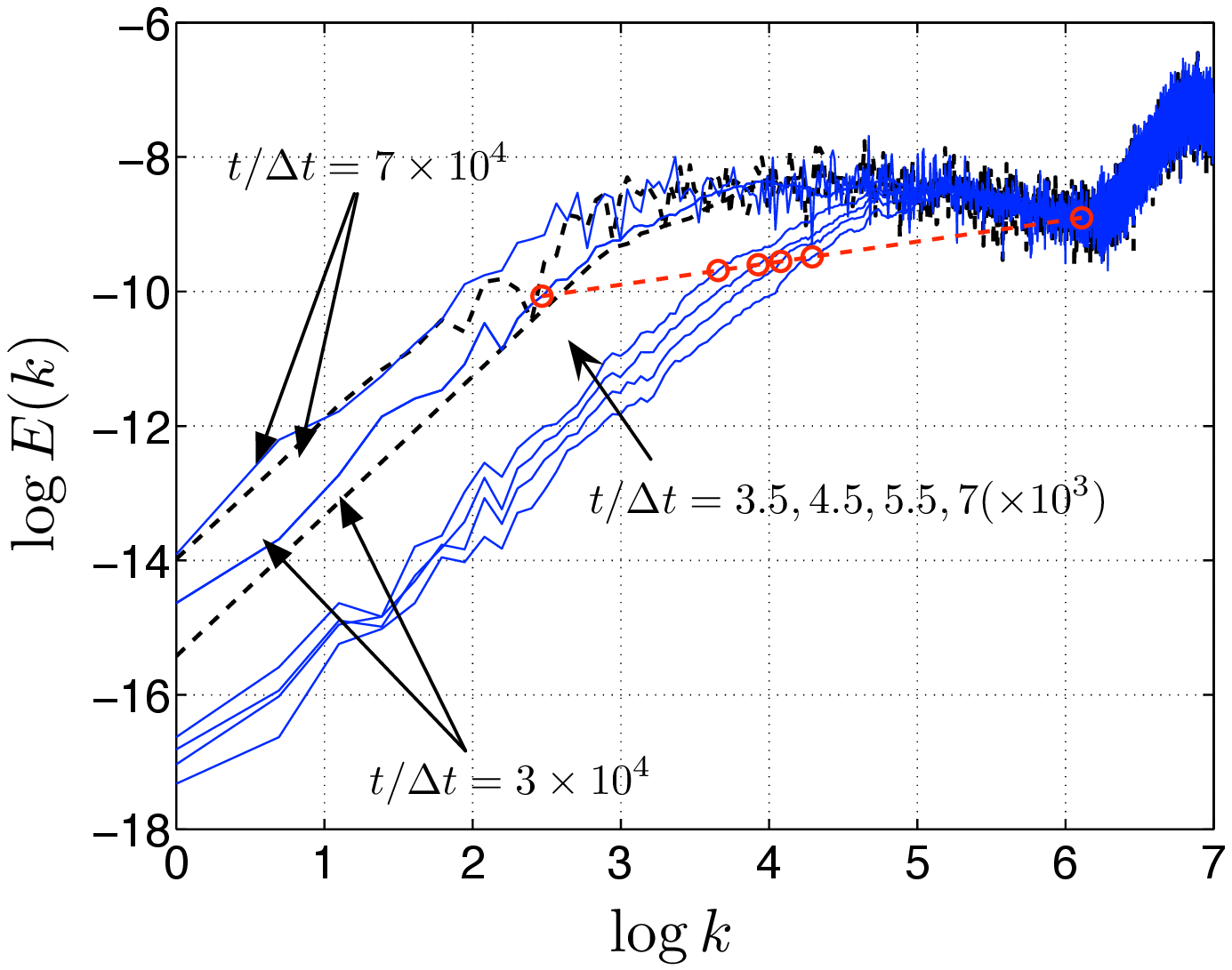} \hspace{0.5cm}
\includegraphics[scale=0.5]{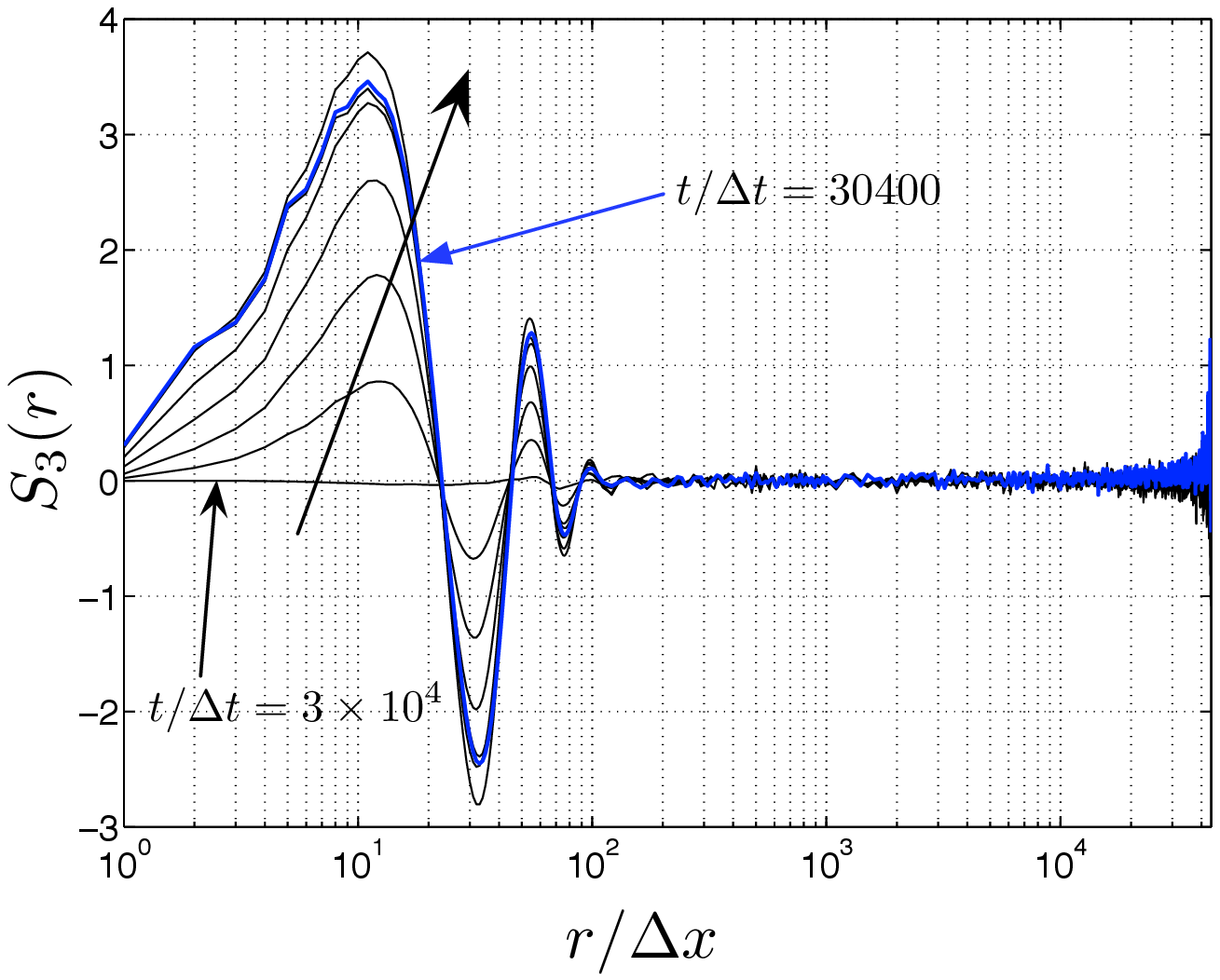}
\caption{(a) Coarse projective integration and its comparison with
the original direct simulation (spectra from the continuing original
simulation are shown in blue).
The phases of all the modes are randomized after projection.
(b) Healing of~$S_3(r)$ (black) in 400~time steps.
After randomizing the phases at~$t=30000 \Delta t$~(flat line),
their transient recovery is depicted after time steps~10, 20, 30,
40, 50, and~400.
Also shown is a comparison with the original simulation at~$t =
30400$ (blue).} \label{fig:strch6} \ec \EFig
Fig.~\ref{fig:strch6} shows an instantiation of such a coarse
projective step.
The restriction of an initial simulation used to estimate the coarse
time derivative are obtained at~$t/\Delta t =$~3500, 4500, 5500,
and~7000; the projection step (which here is performed in {\em
logarithmic} time) brings us to~$t=30000 \Delta t$; the projected
spectrum (black dashed line) is compared to the result of continuing
the full simulation until time~$t = 30000 \Delta t$ (blue).
More importantly, the result of lifting from the projected spectrum
and continuing the evolution till~$t = 70000 \Delta t$ (black dashed
line) is also seen to coincide with the result of a full simulation
till~$t = 70000 \Delta t$ (blue).
Also, the computational savings in the first step of coarse
projective integration can now be quantified.
Using coarse projective integration, the randomly forced Burgers
equation is simulated for 7000~time steps and then a projection step
is used to obtain the spectrum at~$t=30000 \Delta t$, thus resulting
in a saving of 23000~time steps, or a speedup factor of~4.3.
The factor is actually a little lower because of the computational
effort required for the projection step.
Since the dynamics evolve in logarithmic time, this factor will
increase for the later projective integration steps.

The choice of performing projective integration (in effect, Taylor
series expansion of the coarse solution in the independent variable)
in {\em logarithmic} rather than linear time is motivated by the
observation that our transient is close to a dynamically
self-similar solution.
A detailed discussion of the usefulness of of (coarse) projective
integration in problems with continuous symmetries, such as scale
invariance, having (possible approximately) self-similar solutions
can be found in~\cite{mihalis}.

\section{Coarse Dynamic Renormalization}
\label{section:renorm}
Direct simulations starting at~$E(k) = 0$ showed that the
spectrum quickly evolves to a ``steady state" for~$k>k_f \approx
450$; after this initial period, its low wavenumber portion acquires
a ``constant shape'' which is then observed, over time, to stretch
gradually and rather smoothly towards~$k=0$.
This stretching becomes increasingly slower in linear time.
In what follows we do not concern ourselves with the very fast
initial transient (the establishment of a steady state for~$k>k_f$);
nor will we study the {\em very long term} dynamics that occur when
the ``corner" of the spectrum finally approaches~$k=0$.
We focus on the ``intermediate time" regime during which the
$k<k_f$~spectrum acquires its ``constant shape" which then appears
to travel (in logarithmic~$k$ and logarithmic time) towards~$k=0$.
This latter observation suggests an even simpler caricature of the
self-similar evolution; see Fig.~\ref{fig:findconsts}(a): the
spectrum to the left of~$k_f$ acquires quickly the appropriate
``corner like" shape, which we approximate by two straight lines
meeting at a point with wavenumber~$k_0(t)$.
The energy of the wavenumbers to the {\em right} of~$k_0$ remains
more or less unchanged, anchored at the fulcrum point.
The portion of the spectrum to the {\em left} of~$k_0(t)$
subsequently simply moves towards lower~$k$ with constant shape.
A constant shift of this part of the spectrum in~$\log k$
corresponds to a stretching by a constant factor in~$k$, and clearly
suggests that for our apparently self-similar solution
\beq E(k,t) =  E\Big( k \big( t + t_0 \big)^\beta \Big), \qquad
\mbox{for} \quad k<k_0(t), \label{selfsim} \eeq
and an appropriate choice of constants~$\beta$ and~$t_0$.
For the data shown in Fig.~\ref{fig:cartoon1}, initialized with a
zero spectrum at~$t=0$ and taken at times~$t \geq 125 \Delta t$, the
values of the exponent and the constant~$t_0$ are computationally
found to be~$\beta\approx 0.633$ and~$t_0 \approx -98 \Delta t$.
These values can be extracted from Fig.~\ref{fig:findconsts}(b),
which shows plots of the wavenumbers~$k_0(t)$ versus~$t/ \Delta t$
and $k_0(t)$~versus~$(t+t_0)/\Delta t$ on a logarithmic scale.

\BFig \bc \psfrag{d}{\small fulcrum} \psfrag{a}{$k_f$}
\psfrag{n}{\small $k_{0}(t_i)$} \psfrag{m}{\small $k_{0}(t_0)$}
\psfrag{k}{$\log k$} \psfrag{e}{$\log E(k)$} \psfrag{t}{$t/\Delta t,
\, (t+t_0)/\Delta t$} \psfrag{y}{$k_0$}
(a) \hspace{7cm} (b) \\
\includegraphics[scale=0.37]{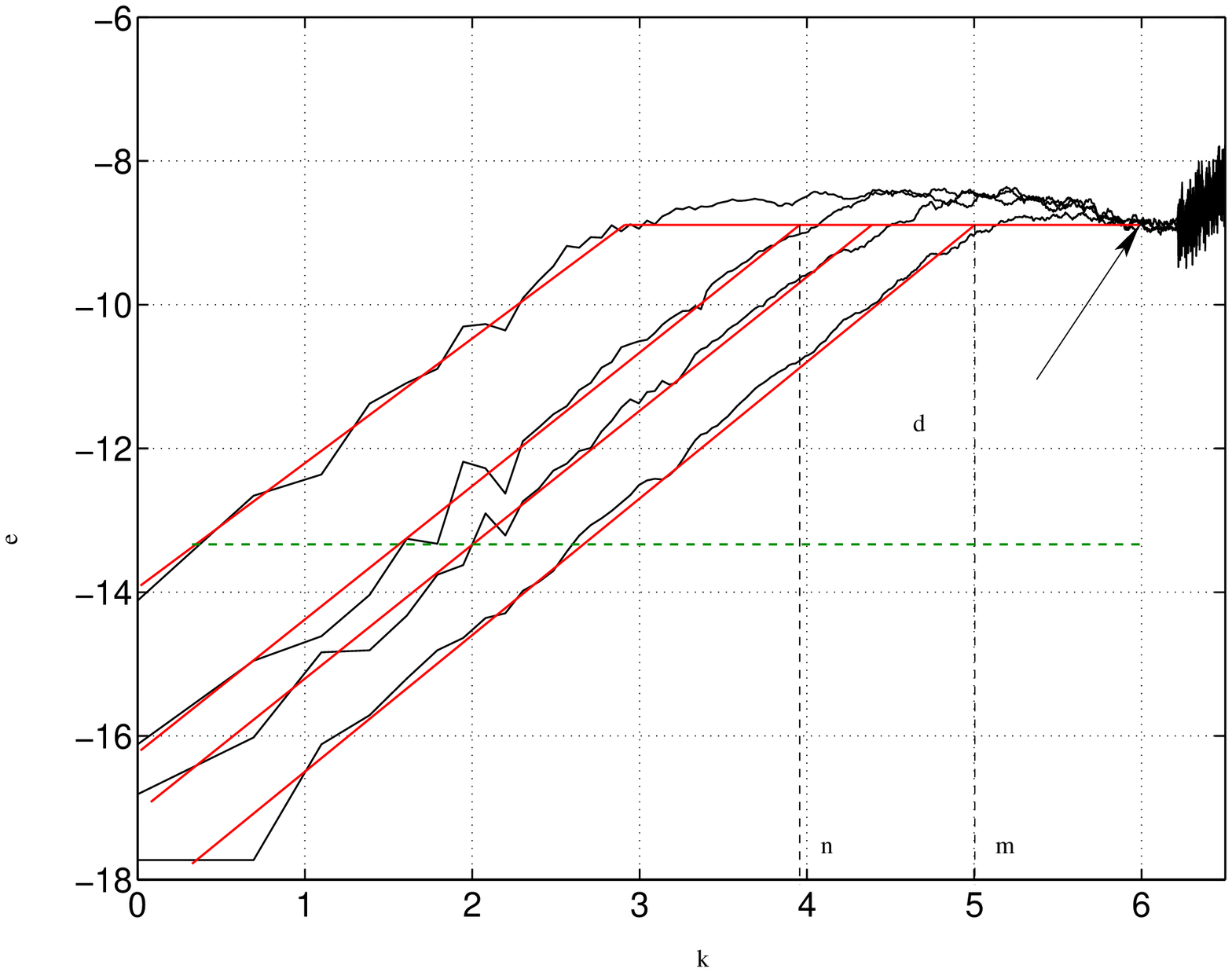}
\includegraphics[scale=0.28]{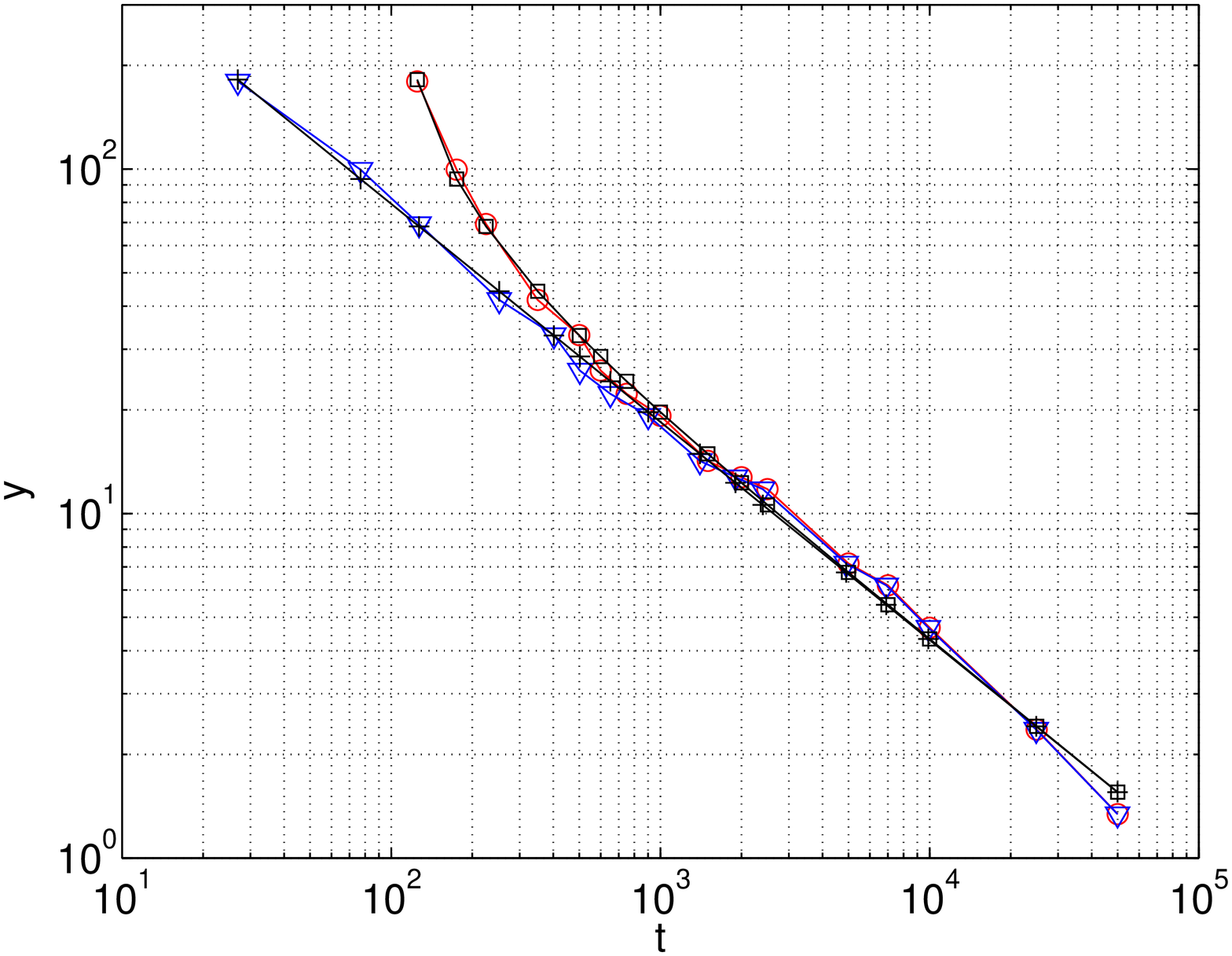}
\caption{ (a) A cartoon illustrating the approximations to the
spectra used to compute the self-similar solution.
The plot shows spectra at a few select time instances.
Their straight line approximations used to compute the exponents are
shown in red.
(b)~Wavenumbers at the intersections of the spectra with the dotted
green line are used to fit the constants~$\beta$ and~$t_0$ (see text).
A plot of these wavenumbers~$k_0$ vs.~$t/\Delta t$ is shown~(red,
$\circ$), with a least squares fit of the form~$k_0(t) =
k_{\mbox{const}} (t + t_0)^\beta$~(black, $\square$).
Also shown are plots of~$k_0$ vs.~$(t+t_0)/\Delta t$~(blue,
$\triangledown$) and the corresponding least squares fit~(black,
+).}
\label{fig:findconsts} \ec \EFig

This appears consistent with the numerical simulations of Yakhot and
She~\cite{vyPRL88} as well as with dynamic renormalization group
results suggesting a non-trivial dynamic scaling of the temporal
frequency~$\omega=O(k^{3/2})$ (or~$kt^{2/3}\approx \mbox{const}$) as
opposed to~$\omega\propto \nu k^{2}$ or~$k=O(\omega^{1/2})$.
The same exponent is obtained using different spectrum
``caricatures'' based on the same low-wavenumber straight-line
approximations (not shown here).
We would like to stress that computation of dynamic scaling
exponents is an extremely difficult and computationally expensive
task.
The fact that the non-trivial exponent~(\ref{exp}) emerged from our
relatively fast calculation seems remarkable.
Moreover, it serves as an independent test of our coarse graining
procedure.

Given these observations, it is clear that the intermediate time
behavior can be recovered in an even more economical fashion than
coarse projective integration: it is enough to find the ``right
shape" (the slope of the leftmost part of the spectrum) as well as
the correct stretching rate (encoded in the exponent); this
information suffices to approximate the spectra at all intermediate
times.
This is analogous to computing the shape and the (constant)
speed of a traveling wave in a problem with translational
invariance.
Knowing the shape and speed we can construct the
solution at any future time; {\em scale invariance} here takes the
place of translational invariance for traveling, and the stretching
rate (quantified by the similarity exponents) is analogous to the
traveling speed.
The problem therefore reduces to finding the ``right" spectrum shape
(for~$k<k_f$): a spectrum~$E(k,t_1)$ which, when evolved for some
time~$T$ to~$E(k,t_1+T)$ and {\em rescaled} in $k$--space, remains
unchanged.
This gives rise to the fixed point problem:
\bea
E(k,\, t_1) - E\Big(k \Big(\frac{t_1 + T + t_0}{t_1 +
t_0}\Big)^\beta, \,t_1 + T\Big) = 0.
\eea
Solving such {\em coarse} fixed point problems using coarse time
steppers has been discussed and illustrated in several
contexts~\cite{chenJNNFM,zouPRE,chenPRE,kesslerPRE}.
The simplest fixed point algorithm is successive substitution: we
start with an initial guess of the self-similar spectrum shape which
we evolve for some time using the direct randomly forced Burgers
simulator, including ensemble averaging.
We then {\em rescale} the resulting spectrum by a constant factor,
and repeat the process; for a discussion of a so-called {\em template
based} approach to choosing the appropriate scaling factor at each
iteration, see~\cite{aronson01,rowley03}, and our choice
for this problem will be discussed below.
If the self-similar spectrum is {\em stable} (as is the case here),
this iteration converges to its scale-invariant shape.
We remark, before proceeding, that other fixed point algorithms (in
particular, matrix-free Newton--Krylov--GMRES,~\cite{kelley04}) have
been used for solving this type of problem; such Newton--based
algorithms can converge to self-similar solutions even if they are
not stable.
In what follows, we demonstrate that successive substitution using a
{\em dynamically rescaled coarse time stepper} does indeed converge
to the right (invariant) spectrum shape.

Long direct simulation will also eventually converge to this
invariant shape; yet, while approaching the right shape, the
transient will also move towards lower wavenumbers (will ``stretch")
and this makes the evolution increasingly slower.
We therefore have the potential to realize a computational advantage
by {\em rescaling} the spectrum after partial evolution to a
constant reference scale, a scale in whose neighborhood the dynamics
are {\em as fast as possible} in physical time; this allows us to
observe the approach to the stable invariant shape {\em without this
shape stretching}.

We now consider another caricature of the (ensemble averaged)
spectra during our intermediate--time simulation, approximated by
two straight lines in the region~$k<k_f$ as shown in
Fig.~\ref{fig:initdiff2}(a).
These two straight lines intersect at a ``corner"; we call this
corner wavenumber~$k=k_c$.
As we discussed earlier in this section, we observe that the slope
of the straight line approximating the region of the spectrum to the
left of the corner remains (approximately) constant~($m=m_{\mbox{left}}$)
during self-similar evolution; finding the right shape becomes then
equivalent to finding this slope.
In order to obtain the self-similar shape, we initialize our
spectrum as follows:
The portion of the spectrum to the right of the fulcrum~$E(k>k_f)$
is initialized at its (already) steady shape.
For~$k_c<k<k_f$, a straight line approximation is used, the slope of
which is computed by a local fitting to the spectrum obtained from
the initial direct simulation.
Finally, for the region~$k<k_c$, we initialize the spectrum with a
straight line of a ``guess'' slope~$m_{\mbox{left}}$.
We now {\em pin} our reference scale by selecting a particular,
fixed~$k_c$; this means that our procedure will converge on the {\em
representative} of the one-parameter family of self-similar shapes
that has its corner fixed there.

Our coarse renormalized time stepper involves {\em lifting} from
this guess spectrum to full Burgers velocity fields, (ensemble)
evolution of these fields by the direct randomly forced Burgers
simulator for a time~$T$, {\em restriction} back to spectra, and
ensemble averaging, after this time.
A further restriction step is now available by our choice of
caricature, in the form of approximating the spectrum by two
piecewise linear segments.
This gives us a {\em new} corner~$k_c$ and a {\em new}
slope~$m_{\mbox{left}}$.
We now {\em rescale} the resulting spectrum by {\em shifting} back
to the old~$k_c$ {\em but keeping the new~$m_{\mbox{left}}$}; this
allows us to persistently observe the system at a convenient scale.
Iterations of this ``lift-evolve-restrict-rescale" procedure is
shown in Fig.~\ref{fig:initdiff2}(a); Fig.~\ref{fig:initdiff2}(b)
shows a sequence of several successive renormalization substitutions
converging to the invariant shape; here it takes approximately six
iterations to convergence if the originally guessed slope is less
steep than the true one; steeper initial slopes practically converge
within one iteration.
\BFig \bc \psfrag{f}{$m_{\mbox{left}}$} \psfrag{e}{\small corners}
\psfrag{d}{\small fulcrum} \psfrag{a}{$k_f$} \psfrag{b}{$k_{c,1}$}
\psfrag{c}{$k_{c,0}$} \psfrag{k}{$\log k$} \psfrag{E}{$\log E(k)$}
\psfrag{E(k)}{$\log E(k)$} \psfrag{x}{$k$} \psfrag{y}{$E(k)$}
(a) \hspace{7cm} (b) \\
\includegraphics[scale=0.42]{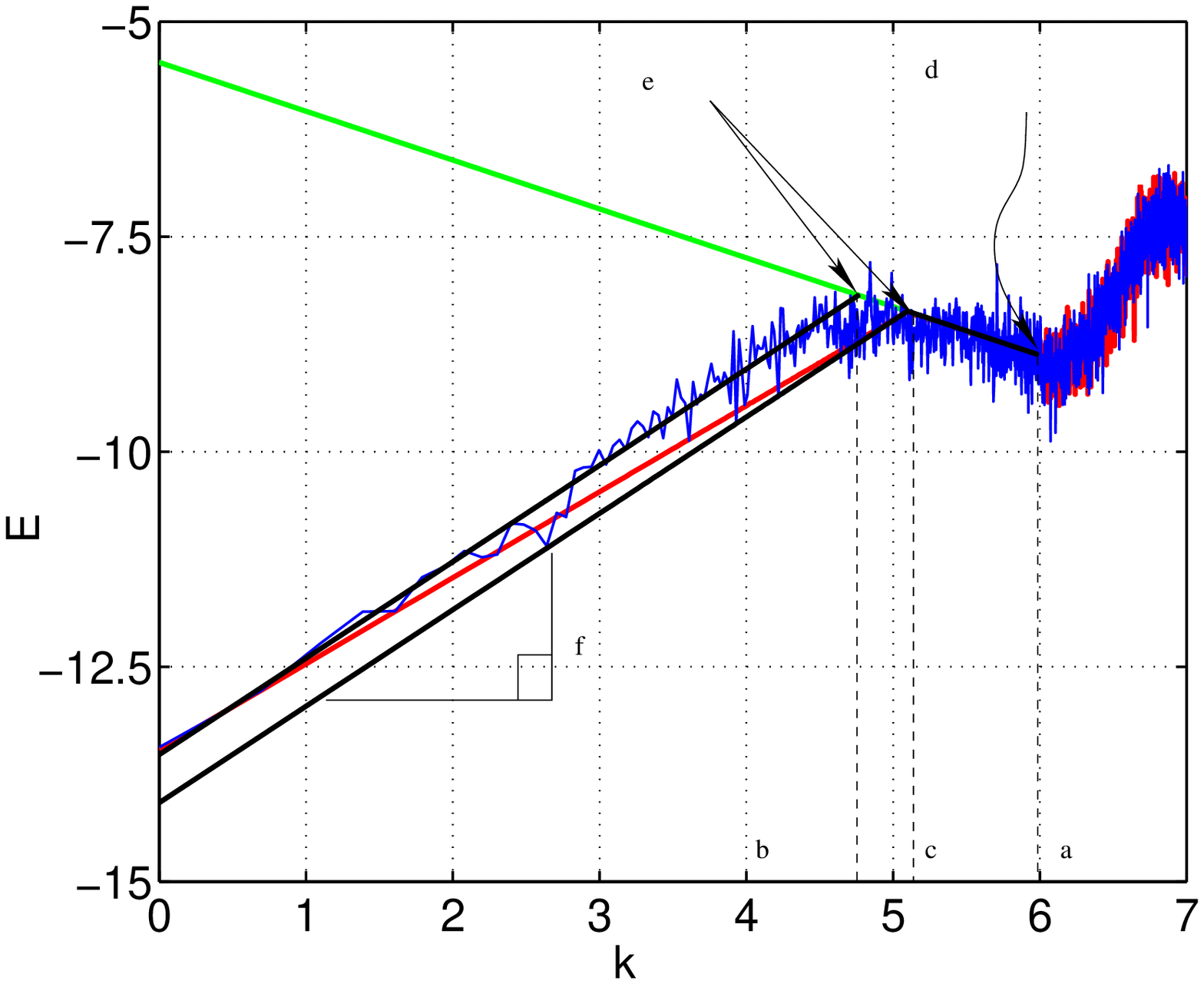}
\includegraphics[scale=0.52]{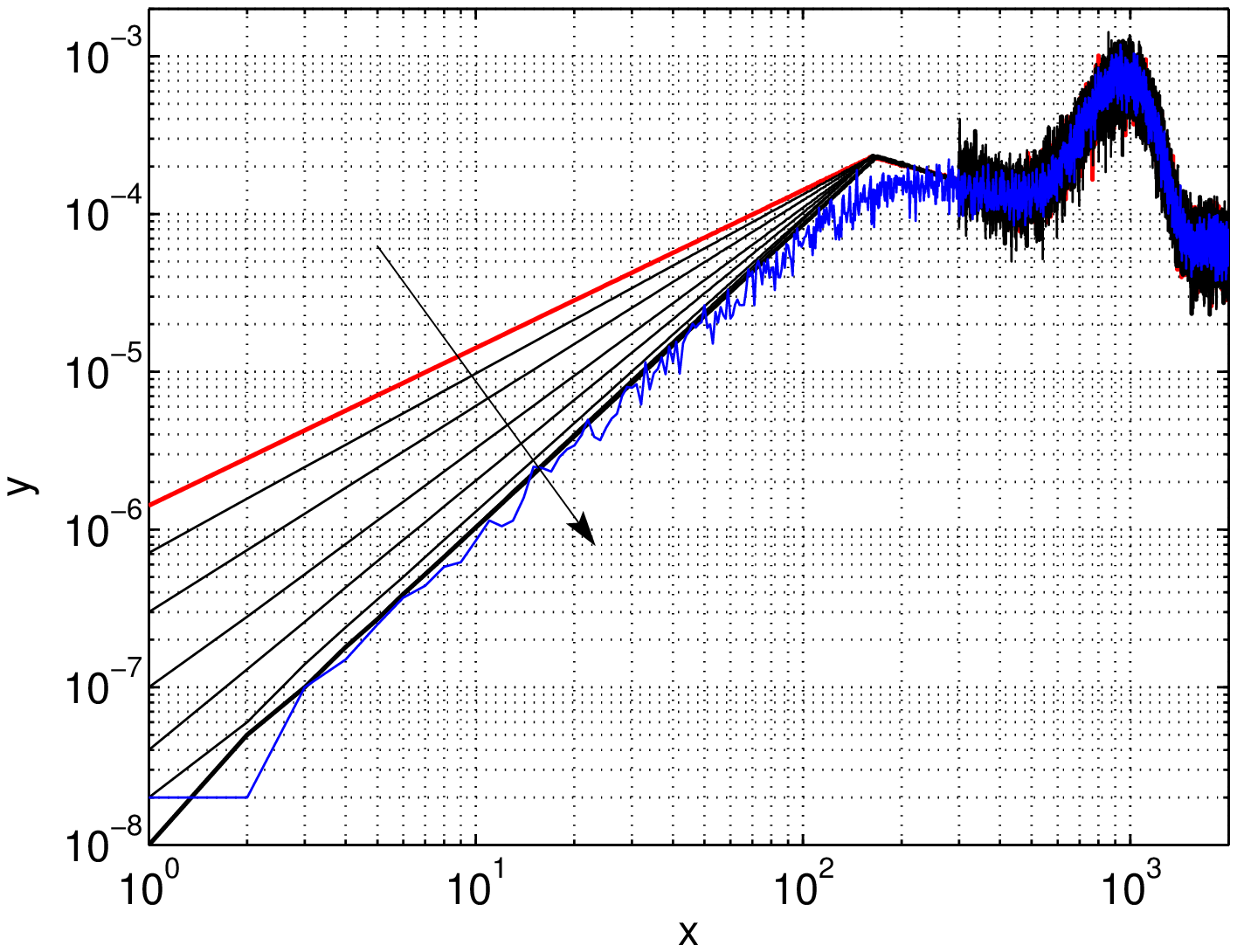}
\caption{ (a) First iteration for the ``tilted-forward'' initial
condition (shown in red).
(b) Sequence of iterations or corrections for the ``tilted-forward''
initial condition.
Final comparison with a representative of the self-similar
transients (shown in blue).} \label{fig:initdiff2} \ec \EFig

From the last iteration (upon convergence) we obtain the shape of
the leftmost part of the spectrum~$E(k,t) \propto k^{1.93\pm 0.05}$
(so~$m_{\mbox{left}} \approx 2$; see~\cite{forster77}).
We can also extract the scaling exponent by doing a similar least
squares fit to a plot of~$\log k_0$~versus~$\log t$.
The constants in the expression~(\ref{selfsim}) are~$\beta = 0.635$
and~$t_0 \approx -6685 \Delta t$ (here we started iterating
at~$t=10000 \Delta t$ of the original simulation).
The larger we can make the corner~$k_c$ (the closer to~$k_f$ we can
practically choose it), the fastest the approach to the self-similar
shape (the dynamics slow down significantly at progressively lower
wavenumbers).

\section{Summary and Discussion}

Using a particular type of randomly forced Burgers equation in one
spatial dimension as our illustrative example, and based on direct
simulations, we postulated that the system dynamics could, under
certain conditions, be coarse grained to an effective evolution
equation for the system energy spectrum.
The direct simulations also strongly suggest an effectively
self-similar evolution of this energy spectrum; it is important to
reiterate that this ansatz is only made for a particular
(intermediate) wavenumber regime, and only for initial conditions
that contain no appreciable energy in the large scales.
Based on these observations we demonstrated the implementation and
use of the ``equation-free" computational methodology: the design of
short bursts of appropriately initialized computational experiments
with the direct simulator that accelerate the study of the system
behavior -- as compared to full direct simulation only.
In particular, we demonstrated coarse projective integration, as
well as coarse dynamic renormalization.
Our computations are predicated upon advance knowledge of the right
set of coarse observables; these are the quantities in terms of
which the (unavailable) evolution equation can {\em in principle} be
written.
Our coarse observables here consisted of (a discretization of) the
(ensemble averaged) energy spectrum; the phase variables were
assumed (and observed, through the use of the third-order structure
function~$S_3$) to become relatively quickly slaved to the energy
spectrum dynamics.
Algorithms that attempt to detect such observables from data-mining
simulation results, such as the diffusion map
approach~\cite{coifman, nadler}, may prove valuable in extending the
applicability of the computational tools we presented here.
The computational construction of synthetic turbulence fields
consistent with such observables (see, for example, the recent
paper~\cite{meneveau06}) can be considered as a type of ``lifting"
scheme in our framework.
Beyond the design of (numerical) experiments to accelerate system
modeling, it is also possible to use such experiments to answer
questions about the nature of the unavailable equation (for example,
whether it is a conservation law).
Our approach made no assumption about the local or nonlocal nature
of the (explicitly unavailable) evolution equation: it may be a
partial differential equation or a nonlocal, integro-differential
one.
It would be particularly interesting to devise computational
experiments with the direct simulation code to explore whether the
right-hand-side of our unavailable evolution equation appears to be
local or not in $k$--wavenumber space.

The method proposed here may prove useful for the efficient
extraction of dynamic scaling exponents in models of various
physical phenomena for which we do not know the governing equations
in closed form.

{\bf Acknowledgements}. This work was partially supported by DARPA
and the~US~DOE(CMPD); the use of computer resources at the Boston
University Scientific Computing and Visualization Group~(SCV) is
gratefully acknowledged.

\end{document}